\documentclass{iaus}
\usepackage{graphicx}

\title[Detecting SNe Ia Progenitors with LISA] %% give here short title %%
{Detecting Double Degenerate Progenitors of SNe Ia with LISA}

\author[Alexander Stroeer, Matthew Benacquista \& Frank Ceballos]   %% give here short author list %%
{Alexander Stroeer, Matthew Benacquista \and Frank Ceballos}

\affiliation{Center for Gravitational Wave Astronomy, University of Texas at Brownsville, \\ 80 Ft. Brown, Brownsville, Texas 78521, USA \\ email: {\tt astroeer@phys.utb.edu}}
\pubyear{2011}
\volume{281}  %% insert here IAU Symposium No.
\pagerange{1--4}
\jname{Binary Paths to Type Ia Supernovae Explosions}
\editors{A.C. Editor, B.D. Editor \& C.E. Editor, eds.}
\begin{document}

\maketitle

\begin{abstract}
The Galactic population of close white dwarf binaries is expected to provide the largest number of gravitational wave sources for low frequency detectors such as the Laser Interferometer Space Antenna (LISA). Current data analysis techniques have demonstrated the capability of resolving on the order of $10^4$ white dwarf binaries from a 2 year observation. Resolved binaries are either at high frequencies or large amplitudes. Such systems are more likely to be high-mass binaries, a subset of which will be progenitors of SNe Ia in the double degenerate scenario. We report on results of a study of the properties of resolved binaries using a population synthesis model of the Galactic white dwarf binaries and a LISA data analysis algorithm using Mock LISA Data Challenge tools.	
\keywords{gravitational waves, binaries: close, white dwarfs}
\end{abstract}

\firstsection % if your document starts with a section,
\section{Introduction}

Space-based gravitational wave detectors, such as the Laser Interferometer Space Antenna (LISA)
\cite[(Danzmann 1996, Hughes 2006)]{danzmann96,hughes06}, will be sensitive to gravitational radiation in the millihertz range. The Galactic population of double white dwarf (DWD) binaries will be the dominant source in this frequency range. With some 30 million systems thought to be producing signals with frequencies above 0.1 mHz, these binaries will produce a confusion-limited signal below $\sim 3$ mHz, with multiple sources in each resolvable frequency bin. Consequently, most individually resolvable signals will come from DWDs that are of high frequency or high mass, or both. The typical double degenerate progenitor of a type Ia supernova consists of two carbon-oxygen white dwarfs in an ultra-compact binary. Many of these are in the class of individually resolvable binaries, and so LISA will be a double degenerate finder that can identify many of the progenitor systems within the Milky Way. In order to estimate any selection biases within the binaries observed with LISA, we have generated a two-year LISA data stream using the DWD population from
\cite[Ruiter \etal~(2010)]{ruiter10}.
We have then applied a detection and parameter estimation algorithm developed by us
\cite[(Stroeer \& Benacquista 2011)]{stroeer11a}
in order to determine the properties of the individually resolved binaries.

\section{Analysis}
We model the population of DWDs for the bulge and the disk of the Galaxy using the {\tt StarTrack} population synthesis code of Belczynski
\cite[(Belczynski \etal~2002, 2008)]{belczynski02,belczynski08}. We assume a constant star formation rate over the past 10 Gyr for the disk and a constant star formation over the first Gyr for the bulge. This results in a population of nearly 30 million DWDs within the LISA sensitivity band. The details of the population synthesis and distribution can be found in
\cite[Ruiter \etal (2010)]{ruiter10}. Using this model, we compute the LISA response to the entire population using the standard LISA configuration of 5 million km armlengths and a two-year observation time. We then compute the integrated signal-to-noise ratio ($\rho$) for every DWD in the population using:
\begin{equation}
\rho^2 = 4\int_0^\infty{\frac{\left|\tilde{h}(f)\right|^2}{S_n(f)}~df}
\end{equation}
where $\tilde{h}$ is the Fourier transform of the expected gravitational wave strain due to a single DWD, and $S_n(f)$ is the one sided instrumental noise spectral density. We select only those DWDs with $\rho \ge 5$ to do a full search using our data analysis pipeline. This first cut gives about 40 thousand binaries. The data analysis pipeline identifies about 31 thousand systems. These are the DWDs we describe as resolved. We define a Type Ia progenitor as any DWD with a total mass above 1.44 $M_\odot$. In the underlying Galactic population, there are 405,309 SNe Ia progenitors out of 27,835,248 DWDs. This ratio drops to 653 out of 39,468 after the first cut, and ends at 503 out of 31,169 resolved systems.

\begin{figure}[t]
%\vspace*{-0.8 cm}
\begin{center}
 \includegraphics[clip=true,trim = 0 0 0 2.0cm,width=0.85\textwidth]{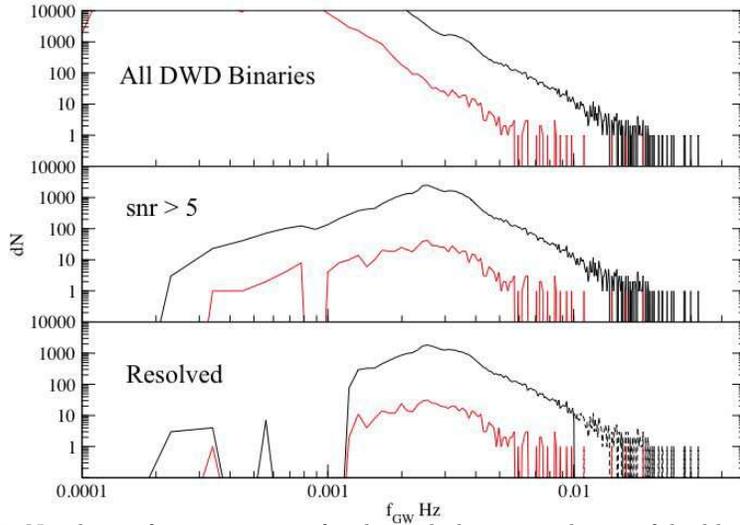}
 \vspace*{-1.0 cm}
 \caption{Number in frequency space for the underlying population of double white dwarfs (top panel), for systems with a signal-to-noise ratio greater than 5 (middle panel), and for systems that were resolved using parameter estimation (bottom panel).}
   \label{fig1}
\end{center}
\end{figure}

The number of resolved systems are heavily weighted toward systems with gravitational wave frequencies above 1 mHz. This can be explained by the fact that there are more than several hundred binaries in each resolvable frequency below 1 mHz, and so systems with lower frequencies are more likely to be lost in the confusion of signals from all the binaries in the same frequency bin. Above this frequency, nearly all systems are recovered, as can be seen in Figure~\ref{fig1}. Thus, the relative number of SNe Ia progenitors compared to the total population can be assumed to be complete at higher frequencies. 

\begin{figure}
\begin{center}
\includegraphics[clip=true,trim= 0 0 0 2.0cm, width=0.85\textwidth]{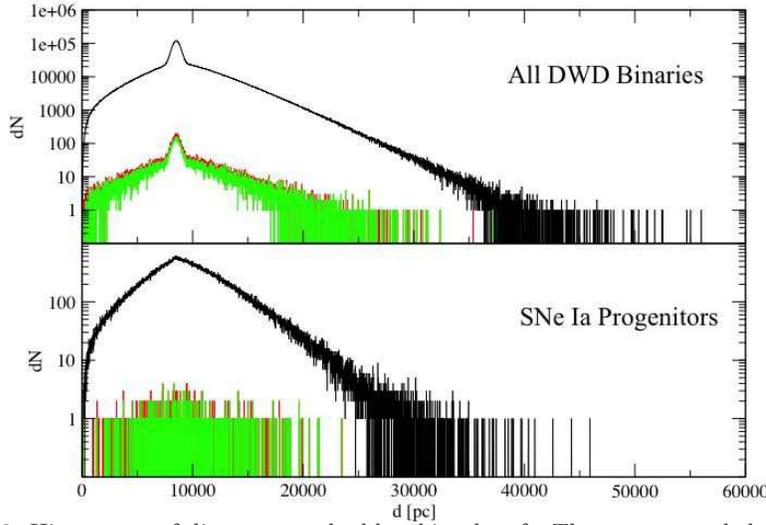}
\vspace*{-1.0 cm}
\caption{Histograms of distances to double white dwarfs. The upper panel shows all binaries while the lower panel is only potential Type Ia progenitors. The underlying population is in black, the systems with a signal-to-noise ratio greater than 5 are in red, and the resolved systems are shown in green.}
 \label{fig2}
 \end{center}
 \end{figure}

When we look at the distribution of distances to the resolved binaries compared to the underlying Galactic population, we find that there is no bias towards nearby systems, as can be seen in Figure~\ref{fig2}. This can be understood by realizing that any binary with an orbital frequency above the confusion-limited cut-off of about 1 mHz will be bright enough in gravitational waves to be visible across the Galaxy. However, as $\left|h\right| \propto d^{-1}$, we do not detect distant systems with $d > 30~{\rm kpc}$.

\begin{figure}
%\vspace*{-2 cm}
\begin{center}
 \includegraphics[clip=true,trim = 0 0 0 2.0cm,width=0.85\textwidth]{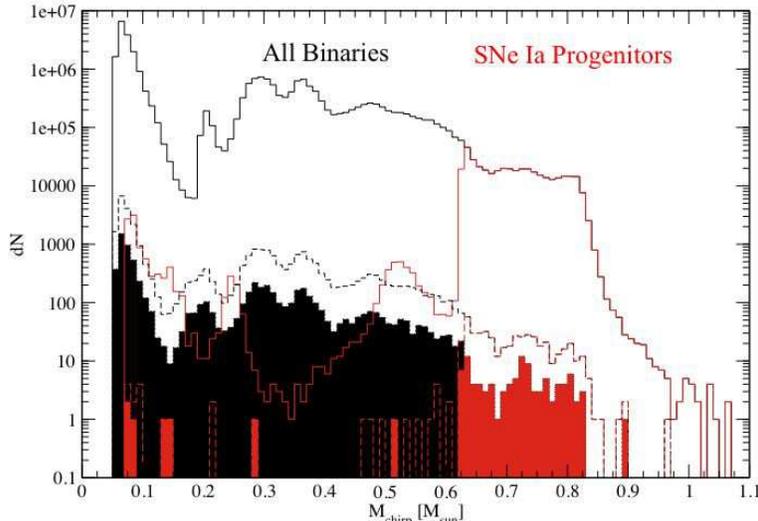} 
 \vspace*{-1.0 cm}
 \caption{Distribution of chirp masses for the underlying population of double white dwarfs and for the population of resolved binaries with measurable chirp (defined as $\dot{f}T_{\rm obs} > 1$). The underlying population is shown in the solid line. The population of all resolved binaries is the dashed line, and the population with measurable chirp is indicated by the filled area. Type Ia progenitors are shown in red.}
   \label{fig3}
\end{center}
\end{figure}

\begin{figure}
\begin{center}
\includegraphics[clip=true, width=0.58\textwidth]{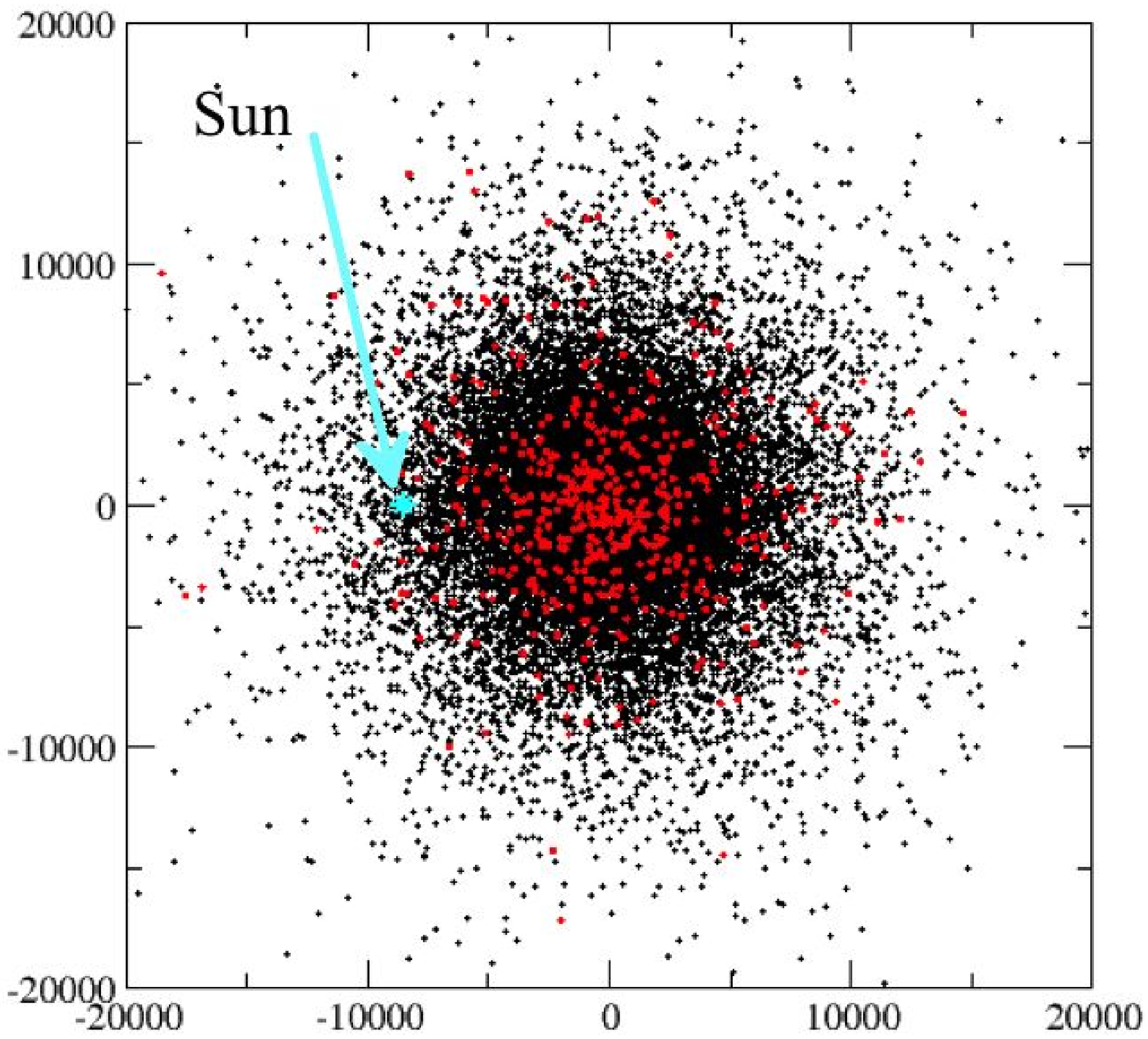}
%\vspace*{-1.0 cm}
\caption{Spatial distribution of resolved binaries viewed from above the Galaxy in galactocentric coordinates. All binaries are shown in black while potential Type Ia progenitors are shown in red. The approximate location of the sun is shown in cyan.}
\label{fig4}
\end{center}
\end{figure}

Parameter estimation in LISA for DWDs returns the amplitude ($\left|h\right|$), the sky location and orientation, and (if measurable) the ``chirp'' or frequency shift due to the emission of gravitational radiation as well as mass transfer and tidal interactions if applicable. In order to identify which resolved binaries are SNe Ia progenitors, we need to determine the total mass. The ``chirp mass'', is given by $M_{\rm chirp} = \mu^{3/5}M^{2/5}$ where $\mu$ is the reduced mass and $M$ is the total mass. If the chirp ($\dot{f}$) can be measured, then $M_{\rm chirp}$ can be found through $M_{\rm chirp} \propto \dot{f}^{3/5}f^{-11/5}$, where the proportionality only depends on physical constants. Once $M_{\rm chirp}$ is measured and the system is ``clean'' so that only gravitational radiation contributes to the chirp, then the distance to the binary can be obtained
\cite[(Schutz 1997)]{schutz97}, 
and so the spatial distribution of these systems throughout the Galaxy will be known. The chirp masses of the resolved binaries are shown in Figure~\ref{fig3}, and it can be seen that all resolved systems with $M_{\rm chirp} \ge 0.62~{\rm M_\odot}$ are SNe Ia progenitors. The spatial distribution of the resolved binaries is shown in Figure~\ref{fig4} and indicates that they are found throughout the Galaxy.

\section{Conclusions}
We have found that LISA will be able to resolve less than 0.1\% of the Galactic DWDs within its sensitivity band of $f \ge 0.1$ mHz. In the underlying population of our synthetic Galaxy, 1.4\% of these DWDs are potential SNe Ia progenitors. The resolved population consists of nearly every binary with a gravitational wave frequency above 3 mHz, and that this population is observable throughout the Galaxy. Type Ia progenitors make up about 1.4\% of the resolved systems, and so there are no appreciable biases in this population. Thus, LISA can be considered to be an effective tool for identifying the Galactic population of DWDs and the SNe Ia progenitors within it.

\end{document}